# Modeling and Evaluating Performance of Routing Operations in Proactive Routing Protocols

D. Mahmood[1], N. Javaid[1], U. Qasim[2], Z. A. Khan[3], A. Khan[1], S. Qurashi[4], A. Memon[5]

[1]COMSATS Institute of Information Technology, Islamabad, Pakistan.
[2]University of Alberta, Alberta, Canada.
[3]Internetworking Program, Faculty of Engg., Dalhousie University, Halifax, Canada.
[4]IT Section, Hamdard University, Islamabad, Pakistan.
[5]IT Section, Associated Press of Pakistan, Islamabad, Pakistan.

**ABSTRACT**

To ensure seamless communication in wireless multi-hop networks, certain classes of routing protocols are defined. This vary paper, is based upon proactive routing protocols for Wireless multi-hop networks. Initially, we discuss Destination Sequence Distance Vector (DSDV), Fish-eye State Routing (FSR) and Optimized Link State Routing (OLSR), precisely followed by mathematical frame work of control overhead regarding proactive natured routing protocols. Finally, extensive simulations are done using NS 2 respecting above mentioned routing protocols covering mobility and scalability issues. Said protocols are compared under mobile and dense environments to conclude our performance analysis.

## I. INTRODUCTION

In this era of infrastructure less communication, wireless multihop networks are gaining popularity day by day. In such networks nodes communicate with each other without any human interface. This makes every node not only act as a transceiver but also a router besides its original functionality. Besides infrastructure less communication provision, such networks gives us liberty of freedom (mobility) and low costs with respect to certain parameters. This concept seems very appealing, as we can apply such kind of networks in almost every field of life. They can be applied in sensor networks, in ad-hoc networks, in body area networks etc. However, in research arena, still there is a lot of work to ensure such freedom and utility. mobile and scalable environments being the prominent aspects have rooms for betterment and to be more efficient.

End to end route surety is responsibility of a routing protocol. Hence protocols of network layer play vital role in smooth, uninterrupted and efficient communication. Protocols dedicated for network layer are solely responsible in establishing/ discovering all required data (w.r.t different routes) from network and than maintaining it.

There are certain classes of protocols defined for network layer of multihop networks. Reactive, Proactive and Hybrid are three major classes of routing protocols having different philosophies of accomplishing the same task. In reactive approach, we tends to use minimum network resources and find a route only when it is required. This defines such protocols as event driven protocols. Network remains idle till the time, a request to find a route is made and when a request is made, such routing protocols at that vary instance start searching route for desired node in network. Considering proactive class of routing protocols, network resources are used initially in a heavy manner in finding every route to any possible destination in network. This philosophy may ensure no tolerance in delay compromising on network resources. The third class, hybrid routing protocols are merger of both reactive and proactive routing protocols [1].

In this work, we are confined only to proactive routing protocols. Initially we discuss three major proactive protocols i.e. DSDV [2], FSR [4] and OLSR [3]. A mathematical model is presented calculating routing overhead in idle and routing overhead in ever varying network. In last section, simulations of above mentioned routing protocols along with extensive comparisons and performance analysis is given.

## II. RELATED WORK

The main objective of routing is efficient energy communication ( [35], [36], [37]). Authors in [6] discuss and present a combined framework of reactive and proactive routing protocols. Their model deals with scalability factor. In [7], authors give analytical model which deals with effect of traffic on control overhead whereas, [8] presents a survey of control overhead of both reactive and proactive protocols. They discuss cost of energy as routing metric. Nadeem *et.al.* [9], enhancing the work of [8], calculate control overhead of FSR, DSDV and OLSR separately in terms of cost of energy as well as cost of time. I.D Aron *et.al* presents link repairing modeling both in local repairing and source to destination repairing along with comparison of routing protocols in [10]. X. Wu *et.al*. [14] give detailed network framework where nodes are mobile and provides *"statistical distribution of topology evolution"*. In [11], authors present brief understanding of scalability issues of network however, impact of topology change was not sufficiently addressed. Authors of [12] and [13] present excellent mathematical network model for proactive routing protocols. We modify the said model by adding control overhead of triggered update messages within the network. Authors of [23] discuss and contribute linear models for proactive routing in wireless multi-hop networks. To examine limitations of presented linear programming models they chose DSDV, FSR and OLSR from proactive routing protocols. Extending this work, [28] presented linear programming for efficient throughput and normalized control overhead. Authors in [24] contribute a path loss model for proactive routing in VANET environment. According to their results, DSDV is most efficient routing protocol under 802.11p. [25] Addresses overall network connectivity and convergence issues for mobile Ad-Hoc and Vehicular Ad-Hoc networks. Security being the key aspect in Ad-hoc or multi-hop networks gains attention by [26]. In this vary paper, authors contributed a secure scheme for wireless proactive routing protocols. [27] Gives a multiple quality of service selection mechanism considering Ad-Hoc networks. A detailed framework of route discovery and route maintenance is produced by [33]. Authors give a generalized model for reactive control overhead. However, in [29] authors contributed generalized routing overhead based on route calculation and route maintenance processes of a proactive routing protocol.
This vary paper is an extension of [29]. In [30], Javaid. N. et al. give improvements in modeling two proactive (FSR and OLSR) and one reactive (DSR) routing protocol. Link duration and Path stability considering DSDV and OLSR is addressed by [31]. There are certain parameters on whome network performance is dependent. Authors of [32] give a detailed analysis on such parameters emphasizing routing protocols of mobile Ad-hoc and Vehicular Ad-hoc networks. In [34] we presented detailed introduction and functionality of DSDV, FSR and OLSR following with mathematical framework on routing load.

## III. PROACTIVE ROUTING

There are many proactive routing protocols in literature (e.g [38], [39], [40], [41]).Whenever a network with proactive routing protocols initiates, route calculation for every possible destination also initiates at that vary time. Contrary to reactive routing in [42], [43], [44], proactive routing doesn't wait for a request to search some destination. On network initialization, each node via flooding grabs global knowledge of network [34]. That knowledge is stored and maintained in form of routing tables routing table at each node regardless of route requirement. That is the reason that latency rate is much lower in proactive routing with respect to reactive routing. However, to achieve low latency rate, a tradeoff is to be made on using network resources. In this study, widely studied and practiced proactive routing protocols as Fisheye State Routing (FSR) [3], Destination Sequenced Distance Vector (DSDV) [2] and Optimized Link State Routing (OLSR) protocols [4] are discussed.

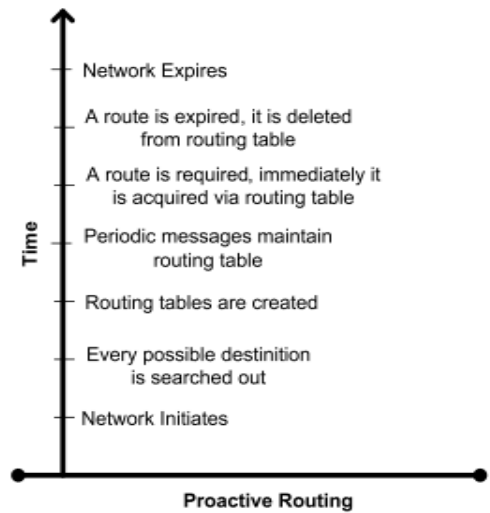

Figure 1: Proactive Routing Approach

### A. Destination Sequenced Distance Vector (DSDV)

DSDV can be claimed as parent routing protocol for Ad-hoc or multi-hop networks. Basically it is derived from Bellman Ford algorithm [17] that provides solutions for shortest path between two nodes [2]. DSDV in addition of classical Bellman Ford Algorithm introduces a new feature i.e. sequence number for each routing table entry of whole of the network. In this protocol, routing table on each node makes lists of all the possible destined nodes within the underlying network along with their number of hops and a sequence number to prioritize the routes. This routing information is broadcasted or multi casted to the neighbors. Besides periodical update messages, DSDV also has a concept of triggered updates whenever a change in topology occurs. C Perkins presented this protocol in $1994$. There is some detailed description of protocol provided in [2] Till to-date, numerous comparisons have been made between DSDV and other routing protocols both reactive and proactive in nature. $Fig.2$ clearly illustrates basic operation of DSDV protocol [34].

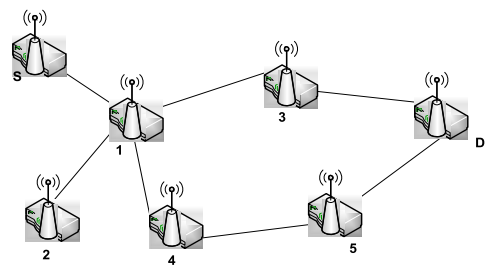

| Routing Table at Node S | | | | | |
|---|---|---|---|---|---|
| Destination | Next hop | Hops | Seq. # | Install time | Metric |
| S | S | 0 | A-846 | 001000 | 0 |
| 1 | 1 | 1 | B-480 | 001200 | 1 |
| 2 | 1 | 2 | C-910 | 001400 | 2 |
| 3 | 1 | 2 | D-540 | 001600 | 2 |
| 4 | 1 | 2 | E-880 | 001600 | 2 |
| 5 | 1 | 3 | F-430 | 001800 | 3 |
| D | 1 | 3 | G-660 | 001800 | 3 |

Figure 2: Routing Table underlying DSDV protocol

*i.* ***Operation: DSDV***

DSDV is table driven routing algorithm and its main feature is to control infinite looping problem by using unique sequence numbers for every packet. Functioning of DSDV is explained in the following algorithm.

1. Initialize Network
2. Procedure link search for all possible destinations
3. Procedure periodic messages broadcast $PM(TR)$
4. $PM(TR) \leftarrow DEST\_ADD, NXT\_HOP$
5. At node $n$ $PM(TR)$ process
6. *if*
7. $PM(TR)\text{ new} == PM(TR)\text{ old}$
8. $SEQ\_NUMold \leftarrow SEQ\_NUMnew$ // replace sequence number
9. *Elseif*
10. $PM(TR) \neq MP(TR)old$
11. $RT \leftarrow PM(TR)new$
12. Flush all $PM(TR\text{ new}$
13. *Elseif*
14. $PM(TR)new == Null$
15. // Link Fail
16. Flush all $TRIGGERED_M SG$
17. $RT \leftarrow updateTRIGGERED\_MSG$
18. // Link Established
19. $BUFFER \leftarrow DATA\_PKT$ till $LINK\_SETTLING\_TIME$
20. Flush all $DATA\_PKT$
21. // Continue Periodic Messages
22. End Procedures

### B.  Fisheye State Routing (FSR)

In proactive routing protocols, converging network using minimal network resources is a huge problem. To address this problem, Fisheye state routing algorithm was introduced giving concept of "Multi level scope". As the correspondent of control or status update messages moves away from the destination, the information propagated gradually declines to lower frequencies. From such status updates or control messages, every node in the network build and maintains a routing table. This routing table is precise for the nodes nearby but as the hop distance increases; the routing information in the same proportion fades or gets imprecise. Hence the route on which a packet travels may seems to be faded but as it gets closer to the destination, route becomes more precise and accurate. FSR follows the link state algorithms as it issue periodical updates of link state, but instead of flooding these periodic updates to whole of the network, it floods in step wise manner. Fisheye state routing is briefly discussed and implemented by [18].

Fish eye state routing is a routing protocol providing a tree like structure. It updates the link state information in different frequencies that depends upon fish eye scope distance. These frequencies are higher for nearer nodes and lower for far away nodes. Within scalable and dynamic environment, a packet as reaches near the destination, the routing gets more accurate regardless of mobility and scalability. FSR basically provide, simplicity of routing, gives updated shortest routes , provide robustness in mobile and scalable environments and one of the major benefits is the reduction of routing overhead [19]. FSR is illustrated in $Fig.3$.

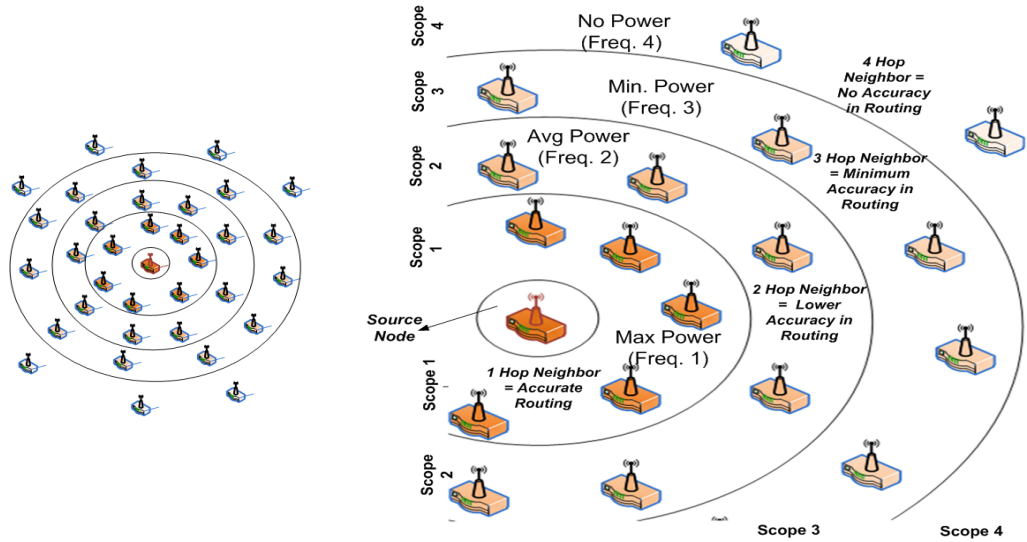

Figure 3: Fisheye Scope

### i. *Operation: FSR*

Fisheye State Routing (FSR) as explained earlier use Fisheye scopes that gives accurate routing for nearby nodes and as distance increases, reliability of routing gradually decreases. The following algorithm shows the working of FSR.

1. // initialize Network
2. Procedures link state packet transmission and reception till $NETWORK\_LIFE\_Time$
3. Procedures Time intervals $TIME[T_{sec}, T_{1sec}, T_{2sec}.....T_{nsec}]$
4. Procedure fisheye scope $FISHEYE[F_1, F_2, F_3....F_n]$
5. For $FISHEYE = 1$, $FISHEYE == F_n$, $FISHEYE^{++}$
6. {
7. For $TIME = 1$, $TIME == T_n$, $TIME^{++}$
8. {
9. Flush all $LS$ //*linkstate*
10. *if*
11. $LS_{new} \neq LS_{old}$
12. Flush all $LS_{new}$
13. $RT \leftarrow LS_{new}$
14. *Elseif*
15. $LS_{new} == null$
16. // until $TIME\_EXPIRES$
17. Flush all $LS_{update}$
18. }

19. }
20. Link Established
21. Flush all $DATA\_PKTs$
22. End Procedures

### C. Optimized Link State Routing (OLSR)

Optimized link state routing protocol is a proactive routing protocol for MANETs [4], [20]. It follows the basic concept of link state routing introducing Multipoint relay concept. Each node in OLSR routing protocol selects a set of multipoint relay nodes which are its neighbors. Only MPRs forward the control packets in such a way that information should reach whole of the network ([21],[22]). These selected MPR nodes are held responsible for declaring LS information in entire network. Multipoint relays are also used in route calculations from a source node to destined node. An MPR selected from a node must have a symmetric or bi directional link to minimize the problems of packet transmissions over asymmetric or unidirectional links. Basically in classical link state routing, only two modifications / optimizations are made to make optimized link state routing protocol [25]. The concept of MPR set of nodes that are responsible of broadcasting topology control messages. And 2ndly contents of topology control message are reduced. Difference between message propagation with and without MPR concept is shown in $Fig.4$

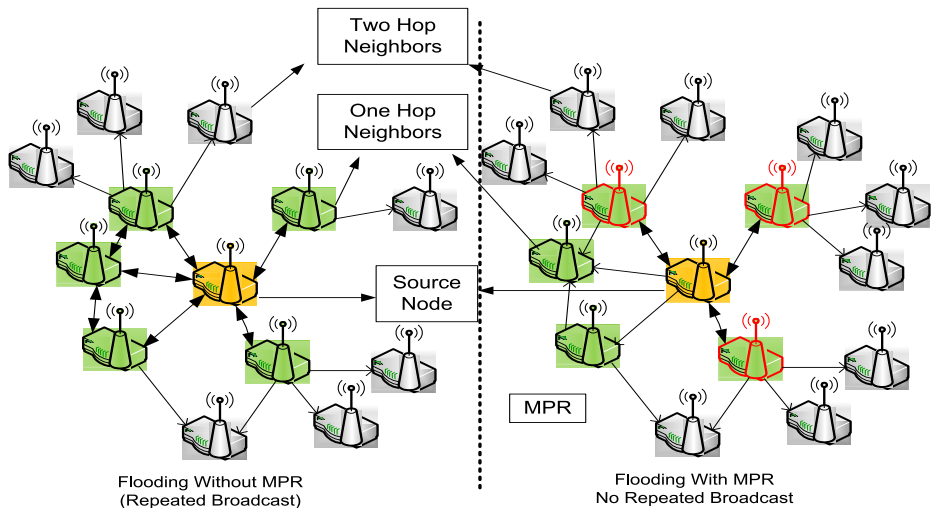

Figure 4: Multi Point Relay Concept

#### i. Operation: OLSR

Optimized link state routing (OLSR) uses the concept of multipoint relays which are responsible for topology control. This concept eliminates overhead to some extant as only MPRs periodically broadcast TC messages not all nodes of a network. Following Algorithm explains the working of OLSR [34].

1. // Initialize Network
2. Procedure Forward HELLO Packet // $FWD\_HELLO\_PKT$
3. Procedure Receive HELLO Packet // $RCV\_HELLO\_PKT$
4. for int $FWD\_HELLO\_PKT = 1$,

$$FWD\_HELLO\_PKT < NETWORK\_LIFE\_TIME,$$
$$FWD\_HELLO\_PKT^+ +$$

5. {
6. if
7. $N\_WILLINGNESS$ 'n' nodes $> N\_WILLINGNESS$ 'm' nodes
8. $MPR \leftarrow$ 'n' nodes
9. }
10. Procedure Forward Topology control packets by 'n' nodes $TC$
11. For int $TC = 1$, $TC > NETWORL\_LIFE\_TIME$, $TC^{++}$
12. {
13. Flush all $TC$
14. }
15. Procedures Compute $TC$ information $TC(i)$
16. *if*
17. $TC(i) \neq TC$;
18. Flush all $TC(i)$
19. *elseif*
20. $TC(i) == TC$
21. $SEQ\_NUMBER\_TC \leftarrow SEQ\_NUMBER\_TC(i)$;
22. continue periodic messages
23. // Link Established
24. End Procedures;

## IV. MATHEMATICAL MODELING

To calculate route discovery and route maintenance routing overhead for proactive routing protocols, we follow the following steps to give enhanced and generalized equations of proactive routing overhead.

1. Network of "N" nodes Initiates
2. Packet failed to reach destination + Periodic updates + triggered updates = = Routing overhead
3. Given in [12] = = number of Packets failed to reach destination
4. Also given in [12] = = periodic update overhead
5. Tpr<T<Tpr+1 = = triggered overhead
6. Number of packets failed to reach destination + periodic update overhead + triggered update overhead = = routing overhead (enhanced equation)
7. Taking parameters of link uptime, periodic update interval, triggered update and packet successfully transferred from equation given in 6
8. Calculate rate of change with respect to above mentioned parameters via partial derivations. (our findings)

## V. MODELING ROUTING OPERATIONS

Control overhead is a vital factor in performance of a routing protocol. In this section, we give a generalized framework considering proactive approach of protocols. Initially we find route calculation overhead, than control overhead generated by dropped packets and finally triggered update messages.

Combining these give us a generalized control overhead of proactive routing protocols.

### A. Proactive Route Calculation Overhead

Considering proactive routing, on initialization of a network, route to each and every possible destinations are created using flooding. In this way a routing table is generated at each node of network. This routing table is kept updated with the help of periodic messages. To ensure very immediate change in network topology, triggered messages are there to cope as discussed in DSDV [2]. Packet loss or drop is another major aspect of control overhead. Packets are dropped due to broken links, change in topology or any radio problem. Normally there are two types of errors that lead to packet failure and are discussed in detail in [9]. In either case, the probability of packet loss is increased.

Considering all this, we can state that, routing overhead for route calculation is sum of number of packets dropped, number of periodic messages and number of triggered update messages. Periodic messages are issued after a specific time period as name indicates to ensure routing table accuracy. Mathematically we can write control overhead as:

$$RO = PF + PR + TR$$

- RO = Routing Overhead
- PF = Packets failed to reach desitnation (dropped packets)
- PR = Periodic messages
- TR = Triggered update messages

### B. Route Failure Impact

Periodic messages are time bound. if any change occurs between two periodic messages, a triggered message is issued. Even than, packet loss happens at this vary time. [12] calculates number of packets failed to reach desitination during periodic message interval ($T_{pr}$) as

$$RO(PF) = (\sum_{p_i \varepsilon PA} \sum_{r=0}^{l_i} Q_r^l(T_{pr}) Na(T_{pr})) \qquad (1)$$

- $RO(PF)$ = routing overhead of packet failures due to link breakage,
- $Q_r^l(T_{pr})$ = probability that during first $r$ hopes, the uplink state does not change its state to down link,
- $Na(T_{pr})$ = number of data packets arriving at time $T_{pr}$ while
- $T_{pr}$ = periodic route update time,
- $L_i$ = length of $Pi$ ($i^{th}$ Path) and
- $PA$ = set of all paths in the network.

### C. Periodic Message Overhead

Major control overhead once after routing tables are established is of periodic messages. These messages are propagated constantly after every specified time period. Control over head in proactive routing can be termed as size of routing table per periodic messages [12].

$$RO(PR) = \frac{kn^3}{BT_{pr}} \qquad (2)$$

- $RO(PR)$ = routing overhead due to periodic updates,
- $B$ = bandwidth,
- $n$ = Number of nodes in a network and
- $K$ = routing protocol impulse factor.

### D. Proactive Route Maintenance overhead

In control overhead besides, packet loss and periodic messages, triggered messages also play a vital role in a network of high mobility.

Suppose a node is mobile in a network and it changes its position between two periodic messages i.e in between $T_{pr}$ and $T_{pr}+1$, say at time $T0$. Routing protocol will not wait for next periodic update to cope routing tables with this change however, to minimize packet loss ratio, protocol issues a trigger message about this change in network topology. This concept is presented graphically for better understanding in fig 5

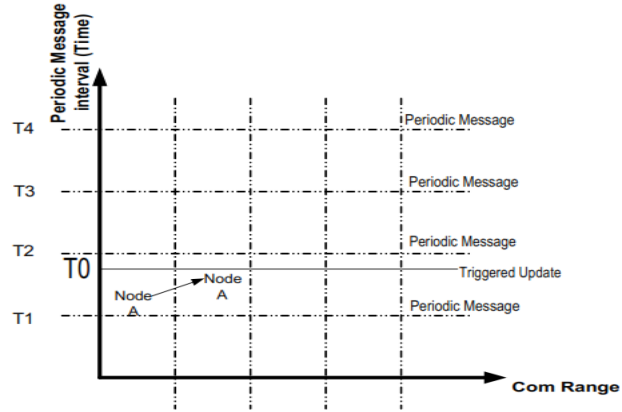

Figure 5: Node A Travels between T1 and T2 from Com Range 1 to 2, Resulting a Trigger update at T0

Fig 5 can also be expressed mathematically as:

$$T_{pr} < T < T_{pr} + 1.$$

As, discussed in [14] this notation can be expressed as:

$$RO(TR)_i = \frac{\left\lceil \dfrac{T}{T_{pr}} \right\rceil}{\dfrac{T}{T_{pr}}} \qquad (3)$$

- $RO(TR)_i$ = routing overhead due to one triggered update message.
- T = Triggered message.
- ceiling operator should be solved by taking the highest possible values (Mathematics Rule).

Considering a very mobile environment, there will be maximum triggered update messages. Eq.3 defines control overhead of one route only, however in VANET or any high mobile environment, over head due to such messages can be calculated as:

$$RO(TR) = \sum_{i=1}^{n} \frac{\left\lceil \frac{T}{T_{pr}} \right\rceil}{\frac{T}{T_{pr}}} \tag{4}$$

- $RO(TR)$ = trigger message overhead
- $T$ = triggered update.

### E. Aggregate Proactive Overhead

Calculating overall control overhead, we combine respective values given in $Eq.1$, $Eq.2$ and $Eq.4$. Eq.5 gives aggregate control overhead of the network.

$$RO = (\sum_{p_i \varepsilon PA} \sum_{r=0}^{l_i} Q_r^l(T_{pr})Na(T_{pr})) + (\frac{kn^3}{BT_{pr}}) + \sum_{i=1}^{n} \frac{\left\lceil \frac{T}{T_{pr}} \right\rceil}{\frac{T}{T_{pr}}} \tag{5}$$

### VI. Modeling Routing Variations

To analyze variations in network with respect to different parameters i.e. periodic update interval, triggered messages, number of nodes of network and uplink time, we define $RO$ given in Eq.5 as an optimized fuction $y$.

$$y(n, T_{pr}, \mu_k, T, \lambda) = (\sum_{p_i \varepsilon PA} \sum_{r=0}^{l_i} Q_r^l(T_{pr})Na(T_{pr}))(\frac{kn^3}{BT_{pr}}) + \sum_{i=1}^{n} \frac{\left\lceil \frac{T}{T_{pr}} \right\rceil}{\frac{T}{T_{pr}}} \tag{6}$$

above mentioned network parameters we chose are represented as:
- $\lambda$ = average number successful packets delivered.
- $\mu_k$ = uplink time
- $T$ = triggered update messages

- $n$ = number of nodes in a network

[7] expresses the probability of initial $r$ hopes where link state do not change i.e. from uplink to down link as

$$Q_r^l(T_{pr}) = 1 - e^{-\frac{rT_{pr}}{\mu_k}} \quad (7)$$

Placing value of $Q_r^l(T_{pr})$ in Eq.6 we get,

$$y(n, T_{pr}, \mu_k, T, \lambda) = (\sum_{p_i \varepsilon PA} \sum_{r=0}^{l_i} 1 - e^{-\frac{rT_{pr}}{\mu_k}} \lambda(T_{pr})) + (\frac{kn^3}{BT_{pr}}) + \sum_{i=1}^{n} \frac{\left\lceil \frac{T}{T_{pr}} \right\rceil}{\frac{T}{T_{pr}}} \quad (8)$$

To calculate variation in these parameters, we take partial derivative of function $y$

### A. Variation in periodic interval time

To calculate variations in routing load of a network with rate of change in periodic update interval time we take partial derivative of function $y$ w.r.t $T_{pr}$

$$\partial y / \partial T_{pr} = (C) \sum_{r=0}^{L_{avg}} 1 - e^{-\frac{rT_{pr}}{\mu_k}} + \frac{rT_{pr}}{\mu_k} e^{-\frac{rT_{pr}}{\mu_k}} -$$

$$\frac{kn^3}{b(T_{pr})^2} + \sum_{i=1}^{n} \frac{\left\lceil \frac{-T}{(T_{pr})^2} \right\rceil + \frac{T}{(T_{pr})^2}}{T^2/(T_{pr})^2} \quad (9)$$

where
- $C = PN_{avg}.\lambda$

### B. Variation in successful packet delivery

The partial derivative of over all routing load with respect to the rate successful packet delivered is expressed as:

$$\partial y / \partial \lambda = (PN_{avg}) \sum_{r=0}^{L_{avg}} T_{pr}(1 - e^{-\frac{rT_{pr}}{\mu_k}}) \quad (10)$$

### C. Variations in Triggered updates

In the same way, if we take partial derivative of function $y$ w.r.t $T$, we get rate of change in triggered update messages:

$$\partial y/\partial T = \sum_{r=0}^{n} \frac{\left\lceil \frac{1}{(T_{pr})^2} \right\rceil - \frac{1}{(T_{pr})^2}}{T^2/(T_{pr})^2} \tag{11}$$

### D. Variation in uplink time

To calculate variation in uplink time, we again have to take partial derivative w.r.t uplink time.

$$\partial y/\partial \mu_k = (PN_{avg} \lambda T_{pr}) \sum_{r=0}^{L_{avg}} -\frac{rT_{pr}}{(\mu_k)^2}(e^{\frac{-rT_{pr}}{\mu_k}}) \tag{12}$$

### E. Variation in scalability of network

Scalability factor play a vital role in control overhead. In varying environment, impact of variation in number of nodes over control overhead is given as:

$$\partial y/\partial n = \frac{3kn^2}{BT_{pr}} \tag{13}$$

### F. Discussions: Variations in network parameters

Eq. 13 expresses rate of change in number of nodes of a network. This equation repreasents that control over head is directly proprotional to number of nodes of a network. As network scalability increases, routing load increases in the same manner. An interesting fact is this that if network comprises of only 3 nodes, there would be no effect on control overhead if one node reduces from network. Supposing that mobility and scalability are constants considering $Eq.9$ and $Eq.11$, it can be inffered that $T_{pr}$ and $T$ are dependant on each other. If periodic message interval increases, triggered update messages also increases. Their relation is not linear however, they are directly proportional to each other. To further explore relations between triggered update messages and periodic message interval time, we take total derivative of $T_{pr}$ and $T$ w.r.t function $y$

$$\frac{dy}{dT_{pr}} = \frac{\partial y}{\partial T_{pr}} + \frac{\partial y}{\partial T}(\frac{dT}{dT_{pr}}) \tag{14}$$

$$dy = \frac{\partial y}{\partial T_{pr}}(dT_{pr}) + \frac{\partial y}{\partial T}(dT) \tag{15}$$

Placing values:

$$dy = (C) \sum_{r=0}^{L_{avg}} 1 - e^{-\frac{rT_{pr}}{\mu_k}} + \frac{rT_{pr}}{\mu_k} e^{-\frac{rT_{pr}}{\mu_k}} - \frac{kn^3}{b(T_{pr})^2} +$$

$$\sum_{i=1}^{n} \frac{\lceil \frac{-T}{(T_{pr})^2} \rceil + \frac{T}{(T_{pr})^2}}{T^2/(T_{pr})^2}(dT_{pr}) + \sum_{r=0}^{n} \frac{\lceil \frac{1}{(T_{pr})^2} \rceil - \frac{1}{(T_{pr})^2}}{T^2/(T_{pr})^2}(dT) \tag{16}$$

Considering low mobility or no mobility in network, longer $T_{pr}$ have no major effect on efficiency of protocol besides it helps reducing control overhead of network increasing efficiency of routing protocol. However, if we consider highly mobile environment, if longer $T_{pr}$ is adjusted, it results in increase in triggered update messages. $Eq.16$ represents that $T_{pr}$ and $T$ have a nonlinear relationship with one another.

If we analyze $Eq.12$ and $Eq.10$, we will come to know that if uplink time equals network life time or if there is $0$ periodic interval time the result of rate of change with respect to uplink time and periodic interval time will be zero. [12]. in other words, if $\mu k$ tends to infinity and $T_{pr}$ is zero, both partial derivatives with respect to $\lambda$ and $\mu k$ will be zero. Assuming, $\partial y/\partial T_{pr} = 0s$, we get:

$$C\sum_{r=0}^{L_{avg}}(1-e^{\frac{-rT_{pr}}{\mu_k}}) + e^{\frac{-rT_{pr}}{\mu_k}} = \frac{kn^3}{(T_{pr})^2} - \frac{\lceil \frac{-T}{(T_{pr})^2} \rceil + \frac{T}{(T_{pr})^2}}{T^2/(T_{pr})^2} \tag{17}$$

The ratio between periodic message interval and uplink time can be termed as an update coefficient [5]. Let us denote that update coefficient as $h = T_{pr}/\mu k$ or $T_{pr} = \mu_k * h$. substituting values of update coefficient gives us optimized control overhead analytical model.

$$C\sum_{r=0}^{L_{avg}}(1-e^{-rh}) + (r*h)^{-rh} = \frac{kn^3}{B(\mu_k * h)^2} - \sum_{i=1}^{n} \frac{-\lceil \frac{T}{(\mu_k * h)^2} \rceil + T(\mu_k * h)^2}{\frac{T^2}{(\mu_k * h)^2}} \tag{18}$$

$Eq.18$ shows direct relationship between update coefficient($h$) and average time for up link. If one of them increases, other also increase however, this relationship is again nonlinear. As inferred from $Eq.13, Eq.18$ also depicts the same relationship between control over head and number of nodes of network.

Considering RFC 3626, there are four periodic messages in $OLSR$ i.e. $HELLO$ messages, Topology Control messages ($TC$), Multiple Interface Declaration messages ($MID$) and host and Network Association messages ($HNA$). In general only $HELLO$ and $TC$ messages are taken into considerations. Understanding basic theme of $OLSR$ routing protocol, Hello messages are propagated for two purposes i.e. knowledge of neighbor hood and selecting an Multi point relay (MPR) set of nodes. This MPR set of nodes is solely responsible for broadcasting received message. The other periodic message i.e. topology control (TC) messages is issued only by MPR set of nodes. According to [3] $HELLO$ message interval is 1 sec while TC message interval should be 2 sec. It clearly states that, TC message time interval must be taken double than Hello message time interval.

Applying the values of $HELLO$ and $TC$ message in routing load optimized model presented in $Eq.18$, we get:

$$RO_{OLSR} = (PN_{avg})\sum_{r=0}^{L_i}(1-e^{-\frac{rT_{pr}}{\mu_k}})\lambda(T_{pr}) + \frac{kn^3}{B*H} +$$

$$\frac{kn^3}{B*2H} + \sum_{i=1}^{n}\frac{\lceil\frac{T}{H+2H}\rceil}{\frac{T}{H+2H}} \quad (19)$$

$H$ = HELLO message interval,
$2H = TC$ message interval (twice the HELLO message interval).
To analyze the time variation in $HELLO$ and $TC$ interval, we partially derivable $Eq.19$ by $H$:

$$\partial y/\partial H = -\frac{kn^3}{B*H^2} - \frac{kn^3}{B*2H^2} + \sum_{i=1}^{n}\frac{\lceil\frac{-T}{(H+2H^2)}\rceil + \frac{T}{(H+2H)^2}}{\frac{T^2}{(H+2H)^2}} \quad (20)$$

Using the presented model, we now are able to precisely calculate desired control overhead of different parameters individually or collectively.

## VII. SIMULATED RESULTS AND DISCUSSIONS

Simulations of DSDV, OLSR [15] and FSR [16] are performed using NS-2. Our main concern is scalability and mobility factors in WMhNs. Simulation parameters are given below:

*Simulation Parameters*

1. Number of nodes = $50$
2. Bandwidth = $2 Mbps$
3. Packet Size = $512 bytes$
4. Size of network = $1000 m^2$.
5. Simulation setup runs on CBR

Within these parameters, we take the following three metrics.

1. Throughput
2. End to end Delay
3. Normalized routing Load.

### A. Simulation Results

For Proactive experiments, we take FSR, DSDV and OLSR, and simulate these routing protocols with respect to mobility and scalability by taking metrics of throughput, delay and normalized routing load.

### B. Throughput of Proactive Routing Protocols

#### i. Mobility Factor:

DSDV outperforms all selected protocol i.e. FSR and OLSR. Main reason of this result is basic functioning of DSDV protocol, that a packet is sent only on the best possible route due to route settling time. Moreover, un-stabilized routes that have the same sequence number in DSDV routing protocol are also advertised with delay. These features of DSDV results in accurate routing hence, throughput is increased. On the other hand, taking OLSR into account, its ability to converge declines as the mobility increases, thus results in lower throughput. Though, in static environment, due to MPR mechanism in OLSR, it gives better throughput than FSR and DSDV. Whenever, a link breaks, there is a concept of triggered messages in DSDV routing protocol that also increase the route accuracy where as in FSR there is no availability of triggered updates. OLSR triggers TC message only when status of MPR's changes.

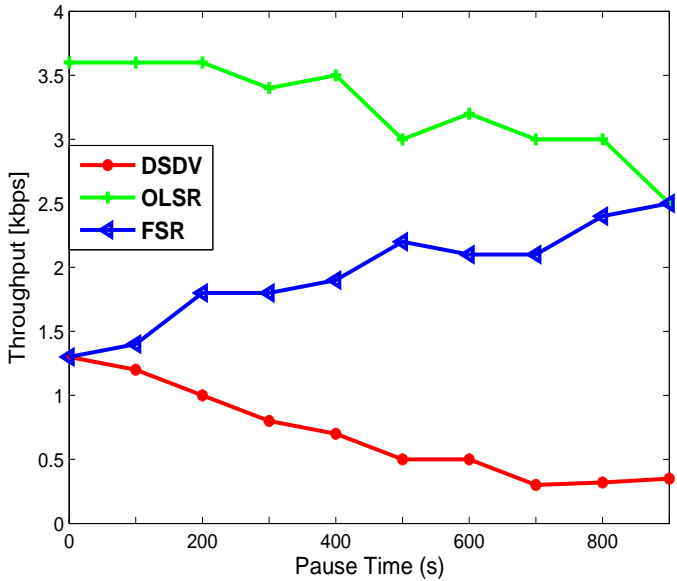

Figure 6: Throughput in Mobile Environment: DSDV, FSR, OLSR

*ii.    Scalability Factor:*

In high scalabilities, OLSR outperforms among chosen protocols. OLSR uses MPR for lowering the routing overhead but periodic messages used to calculate and compute a MPR set for a node take more bandwidth. Though its throughput is more than that of DSDV however. Throughput of FSR also increases as it uses multilevel fisheye scope. This technique results in lower overhead and less consumption of bandwidth which is a major plus point for throughput. DSDV uses Network Protocol Data Units (NPDUs) for lower overhead though, triggered messages create routing overhead, consuming bandwidth and resulting in lower throughput. FSR is highly scalable as it uses different frequencies for different scopes i.e. at different time intervals.

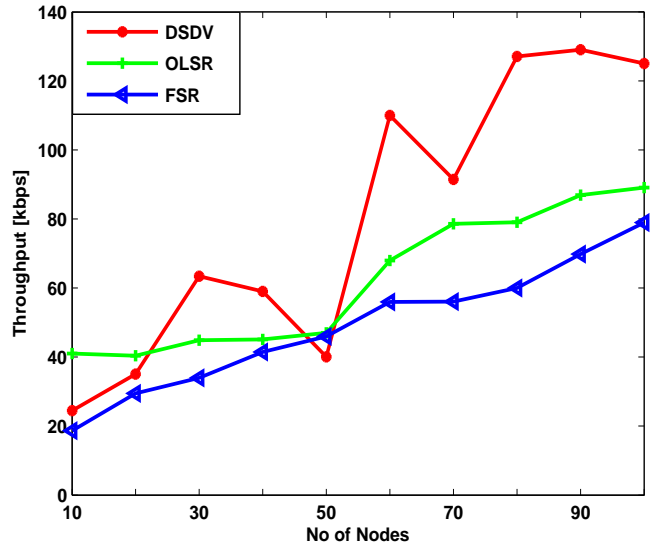

Figure 7: Throughput w.r.t Scalability: DSDV, FSR, OLSR

### C. *End to End Delay of Proactive Routing Protocols*
#### i. *Mobility Factor:*

DSDV proves to be the best for throughput but when considering delay, it bears the worst conditions with respect to FSR and OLSR. Moreover, delayed advertisements of unstable routes results in overall high end to end delay. In DSDV, this is done to reduce the routing overhead and provide route accuracy but it compromises on delay. In such scenario, OLSR performs better than DSDV. FSR produce the highest end to end delay among the studied protocols. As, in the basic theme of FSR, when the mobility increases, the accuracy of far away destined nodes fades. However, as the packet gets closer to destined node, the routing information gets accurate.

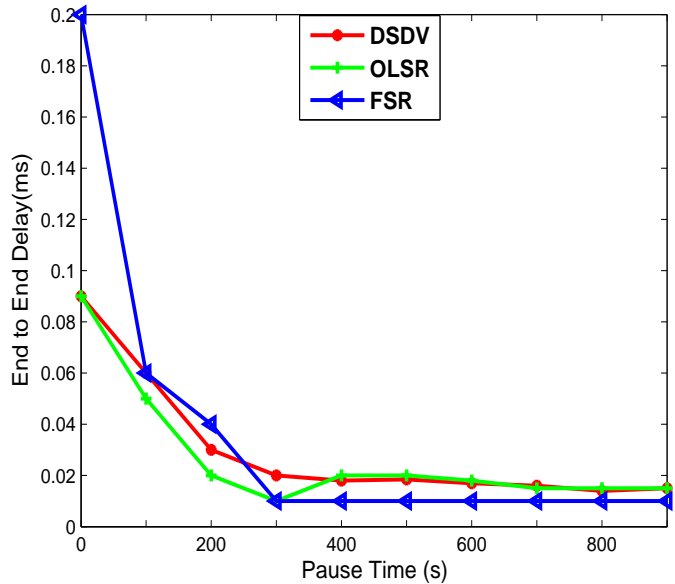

Figure 8: End to End Delay in Mobile Environment: DSDV, FSR, OLSR

### ii. *Scalability Factor:*

As, the network gets dense, end to end delay of discussed routing protocols i.e., FSR, DSDV and OLSR increases. FSR exchanges routing updates with its neighbors in small intervals while information shared at far away nodes has some larger interval. The network become more scalable, end to end delay increases in FSR. In DSDV, end to end delay is due to the two procedures, i.e., finding some routes and then choosing the best route. The network gets denser; end to end also increases. As in proactive nature, the information is spread in whole network. OLSR use MPRs' and in less scalable environment, end to end delay using OLSR is lowered. This is because of MPR concept that presents well organized flooding control instead of flooding a packet on whole network.

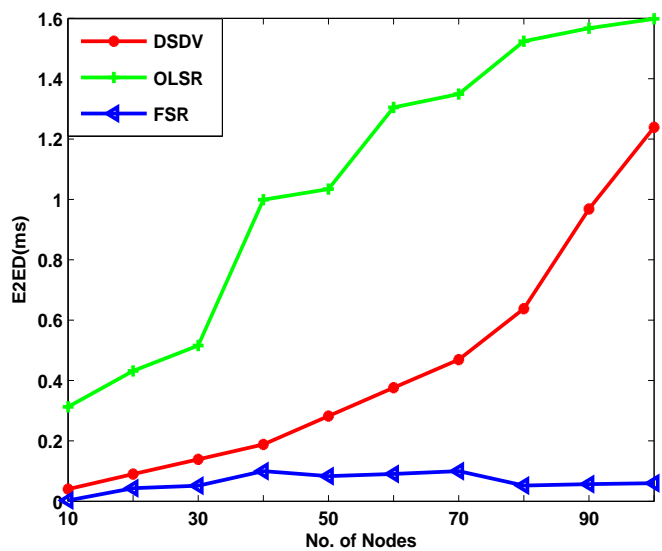

Figure 9: End to End Delay w.r.t Scalability: DSDV, FSR, OLSR

### D. Routing Load of Proactive Routing Protocols
#### i. Mobility Factor:

Among the studied proactive routing protocols, OLSR generates highest routing load due to MPRs computation. DSDV again proves to be a good choice amongst FSR and OLSR in terms of routing overhead. Considering FSR, it bears lower overhead due to control and periodic messages as compared to OLSR. FSR's control messages are periodic based rather event driven based as in OLSR. This feature helps FSR to reduce routing overhead. Moreover, there is limited flooding in FSR i.e., link state information is not flooded among whole network besides, every node manage a link State table which is derived on the basis of up to date information is received. This information is not broadcasted or flooded but is shared amongst neighbors.

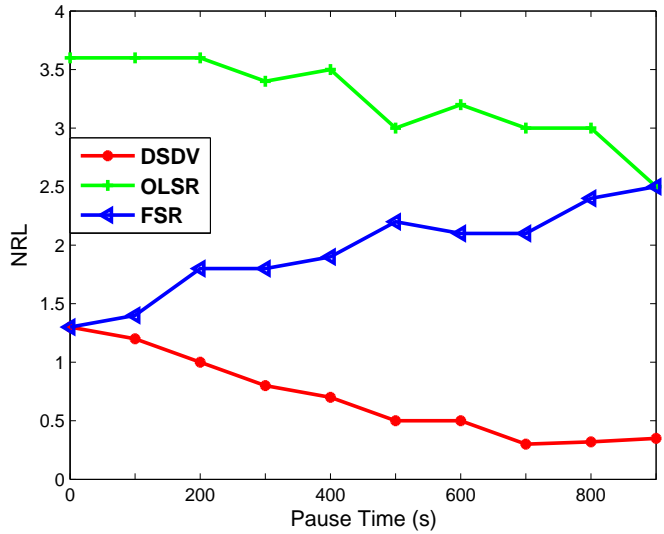

Figure 10: Control Overhead in Mobile Environment: DSDV, FSR, OLSR

#### ii.. Scalability Factor:

OLSR gives the highest routing overhead due to MPR computational messages and $TC$ messages. DSDV and FSR have lower overhead in dense environments. DSDV reduces overhead with the help of NPDUs. The simulated results show that FSR stands best amongst DSDV and OLSR in a dense and mobile environment in terms of overhead.

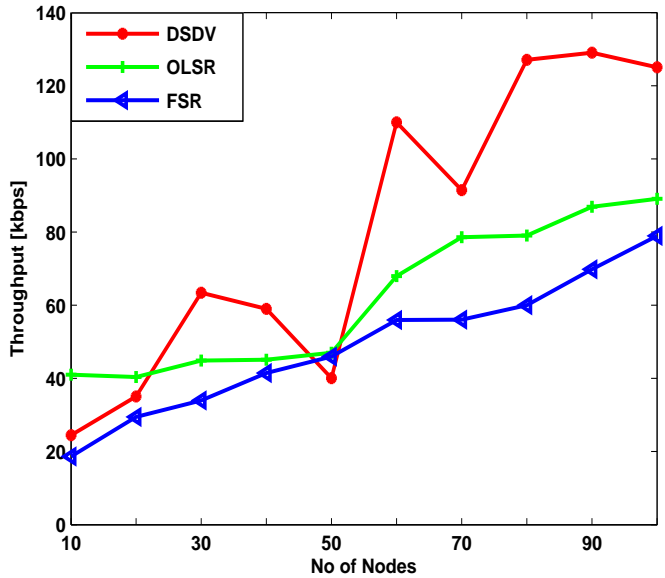

Figure 11: Control Overhead w.r.t.scalability: DSDV, FSR, OLSR

### VIII. Discussion: DSDV Vs FSR Vs OLSR

The protocol that uses minimum resources by its control packets can provide better data flow. Hence, the environments where traffic load is very high, protocols having low routing overhead will survive. If we consider scalability, in proactive routing, OLSR stands tall as it limits retransmissions due to use of MPR concept but only in dense environments. If mobility with the number of nodes of network increases, than FSR is a good choice as it generates low routing overhead that leads to high data rates within the limited bandwidth.

Considering throughput, DSDV proves itself to be the best amongst FSR and OLSR. DSDV sends a packet only on the best possible route which is verified by the protocol twice with a procedure that makes a DSDV route more accurate. This is the reason that DSDV outperforms the rest two routing protocols. OLSR's converging ability minimizes when the environment is mobile else it would prove itself to be the best due to MPR concept.

Considering routing overhead, OLSR is worst due to maximum number of periodic messages for computation of multipoint relays. DSDV proves to be a good choice considering routing overhead as well. Whereas, FSR bears lowest routing load. The feature of Fisheye scope in FSR helps in reducing the routing overhead, as, there is limited flooding i.e., link state information is not flooded among the entire network but is shared with neighbors of a scope only.

#### A. Tabular Representations

Tables presented in our work give a brief performance analysis of studied routing protocols with respect to chosen metrics of throughput, delay and control overhead. The comparisons and performance analysis is result of our experiments performed that are represented graphically as well in previous sections. Table 1 describes basic features, techniques used and distinguish amongst said protocols. Table 2 provides comparison in between DSDV, FSR and OLSR emphasizing mobility factor only. DSDV and FSR gives good results in throughput if mobility is considered, however, they bear higher delay.

Considering Table 3, we give a comparison amongst studied protocols as average performance, good performance and best performance (w.r.t. throughput, delay and control overhead) with respect to different mobility scenarios. Scalability in $DSDV, FSR$ and $OLSR$ is given in Table 4. According to our results, DSDV is better in such an environment where delay is not a big issue. FSR and OLSR respectively are to be used in environment where routing load is not an issue and can be compromised over minimum routing delay.

Table 1: Basic Features: Proactive Routing Protocols

| Feature | FSR | OLSR | DSDV |
|---|---|---|---|
| Protocol Type | Link State | Link State | Distance Vector |
| Route Maintained in | Routing table | Routing Table | Routing Table |
| Multiple Route Discovery | Yes | Yes | Yes |
| Multicast | Yes | Yes | Yes |
| Periodic Broadcast | Yes | Yes | Yes |
| Topology Information | Reduced Topology | Full Topology | Full Topology |
| Update Destination | Neighbors | MPR set | Source |
| Broadcast | Local/ limited | Limited by MPR set | Full |
| Reuse of Routing Information | Yes | Yes | Yes |
| Route Selection | Shortest Hop Count | Hop Count | Shortest Hop Count |
| Route Reconfiguration | Link State Mechanism with Sequence Number | Link State Mechanism/ Routing Messages Transmission in Advance | Sequence Number Adopted |
| Route Discovery Packets | Link State Messages | Via Control Message Link Sensing | Via Control Messages |
| Limiting Overhead | Fisheye procedure, Broadcast Limited only to Transmission Range | Concept of MPRs | Concept of Sequence numbers |
| Collision avoidance, Network Congestion | MAC Layer Protocols only | MAC layer Protocols only | MAC Layer Protocols Only |
| Update Information | Only Neighbor Information | 2 Hop Neighbor Information | By Control Messages |

Table 2: Comparison Proactive Protocols w.r.t. Mobility

| Protocol | Routing Tech. | Pro's | Con's |
|---|---|---|---|
| DSDV | Seq. Number with Avg. Settling Time | Better Throughput in high mobility and lower speed | Delay due to Avg. Settling time. |
| FSR | Multi path routing, Fish eye scope, graded frequencies | Good throughput in highly mobile environment w.r.t. low mobility environment | Higher end to end delay at |
| OLSR | MPR Calculation | Low delay, good throughput in low mobile environment | high control overhead |

Table 3: Performance of Proactive Protocols at different Speeds

| Mobility | Protocol Performing | Delay | Routing Load | Throughput |
|---|---|---|---|---|
| High Mobility (0-300s) puase Timings | Best | FSR | DSDV | DSDV |
| | Average | DSDV | FSR | OLSR |
| | Worst | OLSR | OLSR | FSR |
| Avg. Mobility (300-700s) Puase Timings | Best | DSDV | DSDV | DSDV |
| | Average | OLSR | FSR | OLSR |
| | Worst | FSR | OLSR | FSR |
| Low Mobility (700-900s) Pause Timings | Best | DSDV | DSDV | DSDV |
| | Average | OLSR | FSR | OLSR |
| | Worst | FSR | OLSR | FSR |
| Mixed Mobility (0-900s) Pause Timings | Best | DSDV | DSDV | DSDV |
| | Average | FSR | FSR | OLSR |
| | Worst | OLSR | OLSR | FSR |

Table 4: Comparison Proactive Protocols w.r.t. Scalability

| Protocol | Routing Tech. | Pro's | Con's |
|---|---|---|---|
| DSDV | Avg. Settling Time, Sequence number | Low Control Overhead, High Throughput | Higher Delay |
| FSR | Fisheye Scopes, GF Technique | Lower Delay | Higher Routing Load |
| OLSR | MPR Mechanism | Low Delay | Higher Routing Load |

## XI. Conclusion

Proactive routing protocols i.e. DSDV, FSR and OLSR are studied precisely along with their comparisons and performance analysis with respect to mobility and scalability scenarios. Our study suggests that considering highly scalable environment, OLSR is best option while if maximum throughput is required than DSDV stands best amongs studied routing protocols. To preserve network resources, FSR is a better choice amongst DSDV and OLSR. Besides detailed performance analysis, we modeled routing overhead of proactive natured routing protocols. Aggregate control overhead is further computed to find variations in different network parameters such as scalability (number of nodes), mobility (triggered update messages and periodic interval time), packet delivery ratio (throughput) and uplink time of network. Finally a brief discussion is made regarding such variations in network parameters.

1.